\documentclass[12pt]{iopart}
\usepackage{iopams}
\usepackage[utf8]{inputenc}
\usepackage{todonotes}
\usepackage{amssymb}

\graphicspath{{figs/}}

\begin{document}
%\NJP

\title[H$_2$O adsorption on GaP(100): Optical spectroscopy from first principles]{Water adsorption on the P-rich GaP(100) surface: Optical spectroscopy from first principles}

\author{Matthias M. May$^*$, Michiel Sprik}

\address{University of Cambridge, Department of Chemistry, Lensfield Rd, Cambridge CB2 1EW, UK}
\ead{mm2159@cam.ac.uk}

\vspace{10pt}
%\begin{indented}
%\item[]date: \today
%\end{indented}
%\linenumbers
\begin{abstract}

The contact of water with semiconductors typically changes its surface electronic structure by oxidation or corrosion processes. A detailed knowledge -- or even control of -- the surface structure is highly desirable, as it impacts the performance of opto-electronic devices from gas-sensing to energy conversion applications. It is also a prerequisite for density functional theory-based modelling of the electronic structure in contact with an electrolyte. The P-rich GaP(100) surface is extraordinary with respect to its contact with gas-phase water, as it undergoes a surface reordering, but does not oxidise. We investigate the underlying changes of the surface in contact with water by means of theoretically derived reflection anisotropy spectroscopy (RAS). A comparison of our results with experiment reveals that a water-induced hydrogen-rich phase on the surface is compatible with the boundary conditions from experiment, reproducing the optical spectra. We discuss potential reaction paths that comprise a water-enhanced hydrogen mobility on the surface. Our results also show that computational RAS -- required for the interpretation of experimental signatures -- is feasible for GaP in contact with water double layers. Here, RAS is sensitive to surface electric fields, which are an important ingredient of the Helmholtz-layer. This paves the way for future investigations of RAS at the semiconductor--electrolyte interface.

\end{abstract}

\vspace{2pc}
\noindent{\it Keywords}: gallium phosphide, RAS, spectroscopy, water, density functional theory
%\submitto{\NJP}
%\maketitle

\pagebreak

\section{Introduction}

Water is an ubiquitous component of ambient conditions and its contact with semiconductors impacts areas such as corrosion, sensor applications, solar water splitting or electronic surface passivation of opto-electronic devices. The material class of III-V semiconductors, such as gallium phosphide (GaP) and indium phosphide (InP), is a very interesting case for the semiconductor--water interface, both from a fundamental and an application perspective. In applications, III-V's have demonstrated highest solar energy conversion efficiencies, can be used in (hydrogen) sensing applications, and are widely employed in high-performance opto-electronic devices.\cite{May_pec_tandem_2015, Lin_Pd-InGaP_hydrogen_sensor_2003} For fundamental studies, these semiconductors are well-described by density-functional theory (DFT) and the materials are available in highest quality, which allows for a close interaction between theory and experiment, not deteriorated by complex surface structures. This makes them very interesting model systems for electrochemistry.\cite{Meng_band_alignment_InGaN_water_2016, Pham_modelling_interfaces_solar_water_splitting_2017} Due to its bandgap of 2.26\,eV, GaP should in principle be an interesting candidate for single-absorber photoelectrochemical water splitting. However, the formation of an oxide layer on the initially Ga-rich (100) surface in contact with water creates an unfavourable conduction band offset and significantly reduces obtainable photovoltages.\cite{Kaiser_GaP_2012} A better understanding of surface energetics and structure in contact with water might help to overcome this challenge.

\begin{figure}[htb]
 \centering
 \includegraphics[width=.5\textwidth]{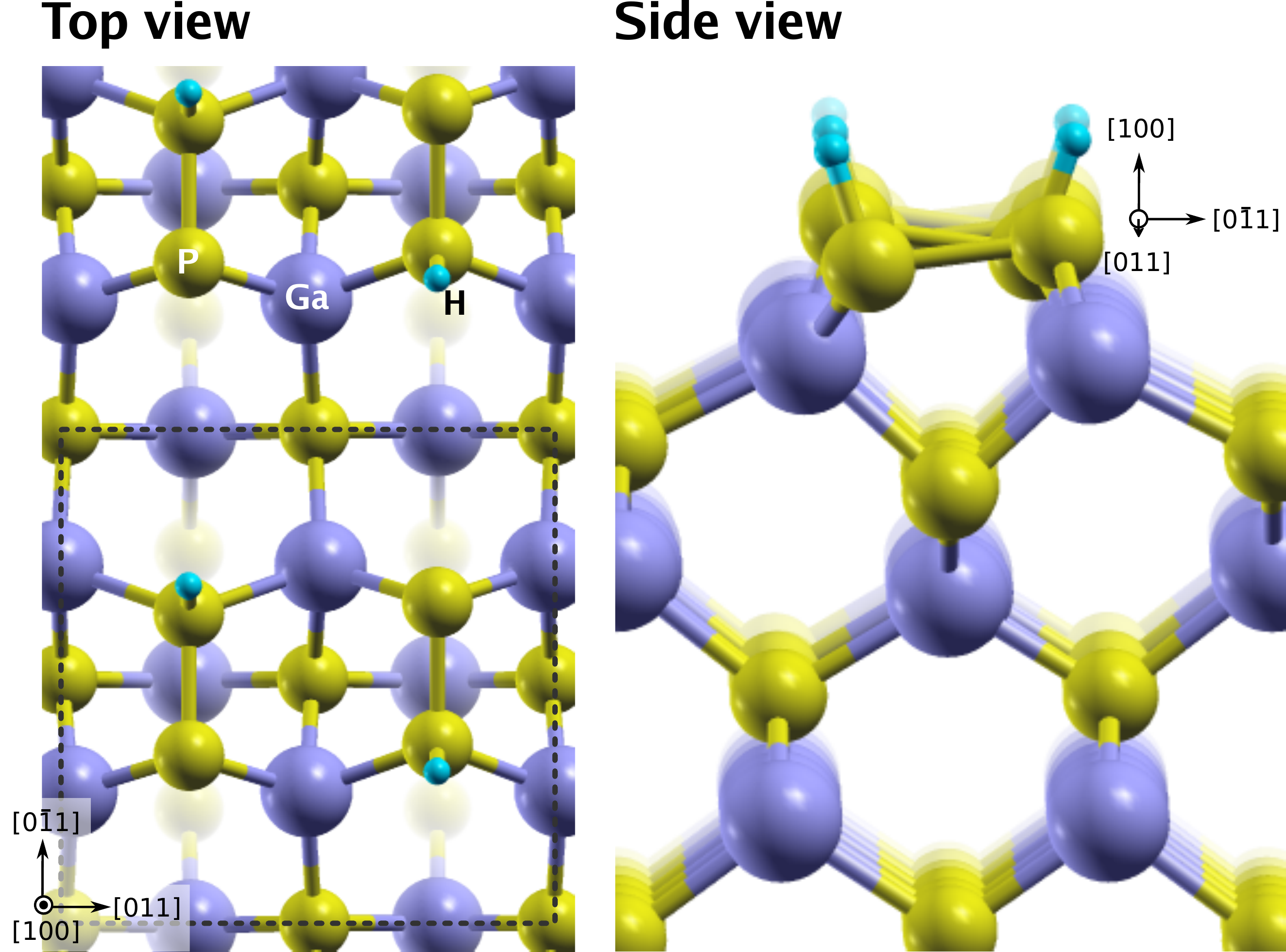}
 \caption{Model of the P-rich, $p(2\times2)$-2D-2H GaP(100) surface (see text for the 2D-2H notation). The dotted square indicates the unit cell of the $p(2\times2)$ reconstruction.}
 \label{fig:P-GaP_struct}
\end{figure}

There are two main surface motifs for polar GaP(100), the Ga-rich, mixed-dimer $(2\times4)$-reconstructed surface and the P-rich, $p(2\times2)/c(4\times2)$ reconstruction. While the former is the more accessible surface, as it can be prepared in vacuum by sputter-annealing, the latter can only be prepared by growth conditions offering hydrogen, due to its P-dimer stabilised by an H atom (see Fig.~\ref{fig:P-GaP_struct}).\cite{Schmidt_InP_hydrogen_stabilized_2003} P-rich reconstructions of (100) III-V surfaces exhibit an extraordinary stability against oxidation under various conditions,\cite{Chen_InP_oxidation_RAS_2002, May_GaP_H2O_2013} which is why we will focus on this surface in the following.

\begin{figure}[htb]
 \centering
 \includegraphics[width=.7\textwidth]{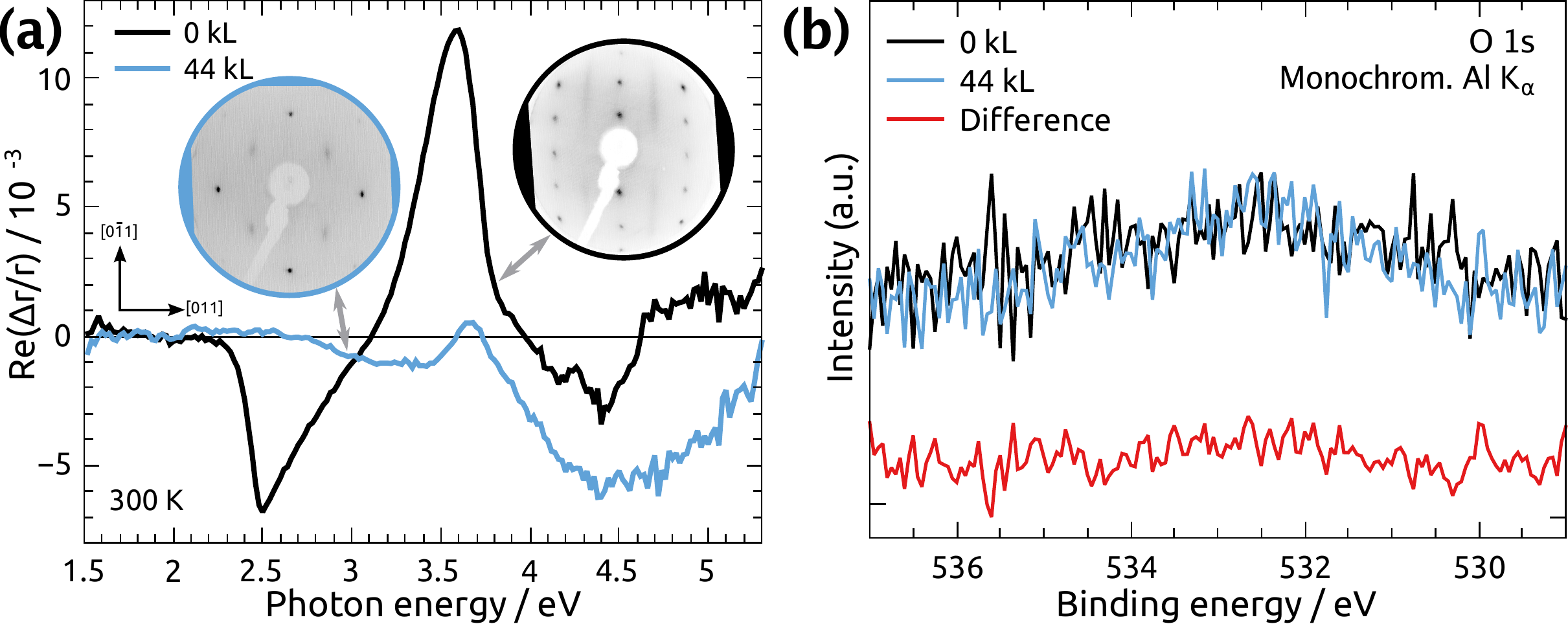}
 \caption{(a)RAS, LEED, and (b) XPS of the surface before (black) and after (blue) water exposure, data from Ref.\cite{May_GaP_H2O_2013}.}
 \label{fig:P-GaP_exp}
\end{figure}

The surface is terminated by phosphorous, which forms buckled dimers that partially saturate their dangling bonds by forming a covalent bond between the downward phosphorous and one hydrogen atom. The mixture of $p(2\times2)$ and $c(4\times2)$-reconstructed surface domains originates in the relative phase of the adjacent P-dimer rows with respect to buckling and the location of the hydrogen atom. In-phase dimer rows result in the $p(2\times2)$ reconstruction depicted in Fig.~\ref{fig:P-GaP_exp}(a), while a phase shift of $\pi$, corresponding to the H atoms on the opposite ends of the dimers, is described by the $c(4\times2)$ reconstruction. Both P-rich motifs are in principle present on real surfaces, as they can inter-convert by ``dimer flipping'' at room temperature.\cite{Kleinschmidt_dimer_flipping_2011} Another notation for the surface, that also describes surface stoichiometry with the number of dimers (D), sole phosphorous (P), and hydrogen atoms (H), together with the symmetry is $(2\times2)$-2D-2H.\cite{Letzig_2x2_reconstruction_InP_2006} The exact structure of the P-rich surface first remained elusive, but was finally resolved by a combination of theory and experiment, employing surface-sensitive optical spectroscopy.\cite{Schmidt_InP_hydrogen_stabilized_2003}

Gallium phosphide and its interaction with water has been subject of a number of studies, also because it is one binary extremum of the ternary Ga$_{1-x}$In$_x$P compound (in the following GaInP), a typical absorber material in III-V solar cells.\cite{Kaiser_GaP_2012, May_GaP_H2O_2013, Wood_hydrogen-bond_dynamics_2013, Wood_surface_chemistry_water_InP_GaP_2014, Kronawitter_neg_charged_hydride_hydroxide_GaP_H2O_2015, Zhang_GaP_water_2016} The polar (100) surfaces, which we will discuss in the following, are currently the most relevant for solar energy conversion applications.

Exposure of the pristine Ga-rich GaP(100) surface to water vapour was both modelled by DFT and investigated experimentally, finding dissociative adsorption of water, leading to a hydroxylised surface.\cite{May_GaP_H2O_2013, Wood_surface_chemistry_water_InP_GaP_2014} The dynamics on the III-rich surface in contact with bulk water, modelled by Car-Parrinello molecular dynamics, suggests a high proton mobility via local hopping on the surface for the closely related InP(100), but not for GaP(100).\cite{Wood_hydrogen-bond_dynamics_2013}

Data on the P-rich surface in contact with water is, however, limited so far. Clean, atomically well-defined surfaces were prepared by metal-organic vapour phase epitaxy (MOVPE) under optical \textit{in situ} control and transferred contamination-free to ultra-high vacuum, where they were exposed to water vapour at 300\,K.\cite{May_GaP_H2O_2013} They showed a completely different behaviour than the Ga-rich surface: Low-energy electron diffraction (LEED) revealed a transition to a new $c(2\times2)$ surface reconstruction after water exposure (Fig.~\ref{fig:P-GaP_exp}a). \textit{In situ} reflection anisotropy spectroscopy (RAS), which probes the optical anisotropy with very high surface sensitivity in the case of cubic (100) surfaces, also indicated a new surface structure as depicted in Figure~\ref{fig:P-GaP_exp}(a). This is in contrast to the Ga-rich surface, where the surface ordering was partially lost and no new superstructure arose. Surprisingly, however, no evidence of oxygen was found by X-ray photoelectron spectroscopy (XPS, Fig.~\ref{fig:P-GaP_exp}b), even after high exposures in the order of 50\,kL. Reaction pathway and the resulting surface structure remained elusive, yet knowledge of the latter is required as a starting point for models of the energetics at the surface in contact with an aqueous electrolyte.

As a probe to study surfaces and interfaces under real conditions, surface-sensitive optical spectroscopy is highly desirable, yet only very few methods with the required spatial resolution are applicable in growth environment or electrochemistry.\cite{Esposito_methods_spatial_photoelectrode_characterisation_2015,Supplie_admi_review_2017} Reflection anisotropy spectroscopy is such an optical technique, probing the anisotropy of a surface at very high surface sensitivity,\cite{Aspnes_Studna_RAS_1985} and with additional high in-plane resolution in the case of RA microscopy.\cite{Punckt_RAM_2007} Linearly polarized light impinges on a sample at near-normal incidence and the reflected light is discriminated with respect to two orthogonal axes, here (and in the following) the $[011]$ and $[0\bar{1}1]$ directions of a cubic (100) surface:

\begin{equation}
\label{eq:ras}
 \frac{\Delta r}{r}=2\frac{r_{[0\bar{1}1]}-r_{[011]}}{r_{[0\bar{1}1]}+r_{[011]}}~,~r\in\mathbb{C}
\end{equation}

Not limited by the short mean free path of electrons, the typical surface probe in surface science, it is applicable in gas ambient, vacuum, but also at the solid--liquid interface. The latter feature makes it an attractive technique for \textit{in situ} studies of electrochemical systems, where potential-dependent optical signatures could be identified at the interface between Au and an electrolyte.\cite{Weightman_RAS_Au110_liquid_environments_2015} An interpretation of spectral features beyond correlation to other experimental signatures does, however, require modelling of the electronic structure and derivation of optical excitations.\cite{Schmidt_calculation_of_surface_optical_properties_2004}

In this paper, we study the interaction of P-rich GaP(100) with water vapour by means of density-functional theory and theoretically derived reflection anisotropy spectroscopy. Comparing the calculated spectra of energetically favourable surface structures with experiment, we find a hydrogen-rich surface, where two additional H atoms are inserted into the P-dimer, as the most probable surface geometry after water exposure. This enables us to provide a starting structure for further modelling of the initially P-rich GaP(100) interface to water. Furthermore, we discuss the feasibility to model RAS at solid--liquid interfaces.

\section{Computational details}

\subsection*{Ground-state DFT}

Electronic structure calculations of the ground state were first performed with the CP2K/QUICKSTEP code in the Gaussian-And-Plane-Wave scheme with GTH pseudopotentials and a plane-wave cutoff of 500\,Ry.\cite{VandeVondele_CP2k_2005, CP2K_4-1_2016, Goedecker_separable_Gaussian_pseudopotentials_1996} Exchange-correlation was described by the PBE potential, dispersion correction was included at a DFT-D3 level and the Brillouin Zone sampled at the $\Gamma$ point only.\cite{Grimme_DFT-D_dispersion_correction_2010, Perdew_GGA_made_simple_1996} After geometry optimisation in CP2K, where forces where converged to $10^{-5}$\, Ha/Bohr, coordinates were transferred to PWSCF, which is part of the QUANTUM ESPRESSO (QE) distribution.\cite{Gianozzi_quantum_espresso_2009} The lattice constant was adapted to the respective ground state described in a plane-wave framework. In QE, the electronic structure was computed with optimized norm-conserving pseudopotentials\cite{Schlipf_ONCV_pseudopotentials_2015} and a 60\,Ry energy cutoff. A $(7\times7\times1)$ Monkhorst-Pack grid was used for the k-points. Symmetric, tetragonal supercells were composed of 5 atomic double-layers of GaP, containing two $(2\times2)$ unit cells of the surface reconstruction, separated by 21\,\AA{} of vacuum.

\subsection*{Calculation of RAS}

RA spectra were calculated from the QE electronic structure with the many-bode code YAMBO in the random phase approximation.\cite{Marini_yambo_2009} A Gaussian smearing of 100\,meV was used, as well as a rigid shift of 0.71\,eV for the bandgap to compensate for the difference between PBE (1.55\,eV) and experimental (2.26\,eV)\cite{Kaiser_GaP_2012} bandgap. The half-slab of interest was selected by a real-space cutoff and we used an experimental bulk dielectric function.\cite{Hogan_dielectric_function_finite_slabs_2003, Aspnes_dielectric_functions_1983}

\section{Results \& Discussion}

In principle, a density functional theory-based molecular dynamics (DFTMD) simulation starting from the P-rich surface that is exposed to gaseous water should eventually feature the reaction between surface and water molecules, leading to the correct surface (electronic) structure after water exposure.\cite{Pham_modelling_interfaces_solar_water_splitting_2017} Timescales presently accessible to DFTMD are, however, too short to feature the whole reaction, especially if reaction kinetics are slow, as indicated in the case here by the high exposures in the order of 50\,kL required experimentally to reach the final surface state. For the clarification of the surface structure, we therefore took the approach to model several surface geometries compatible with the boundary conditions from experiment, derive RA spectra, and compare them to experimental spectra. The boundary conditions as derived from LEED and XPS \cite{May_GaP_H2O_2013} were i) no oxygen on the surface after water, ii) a $c(2\times 2)$ symmetry, and iii) an optical anisotropy as given in Fig.~\ref{fig:P-GaP_exp}(b).

\begin{figure}[htb]
 \centering
 \includegraphics[width=.7\textwidth]{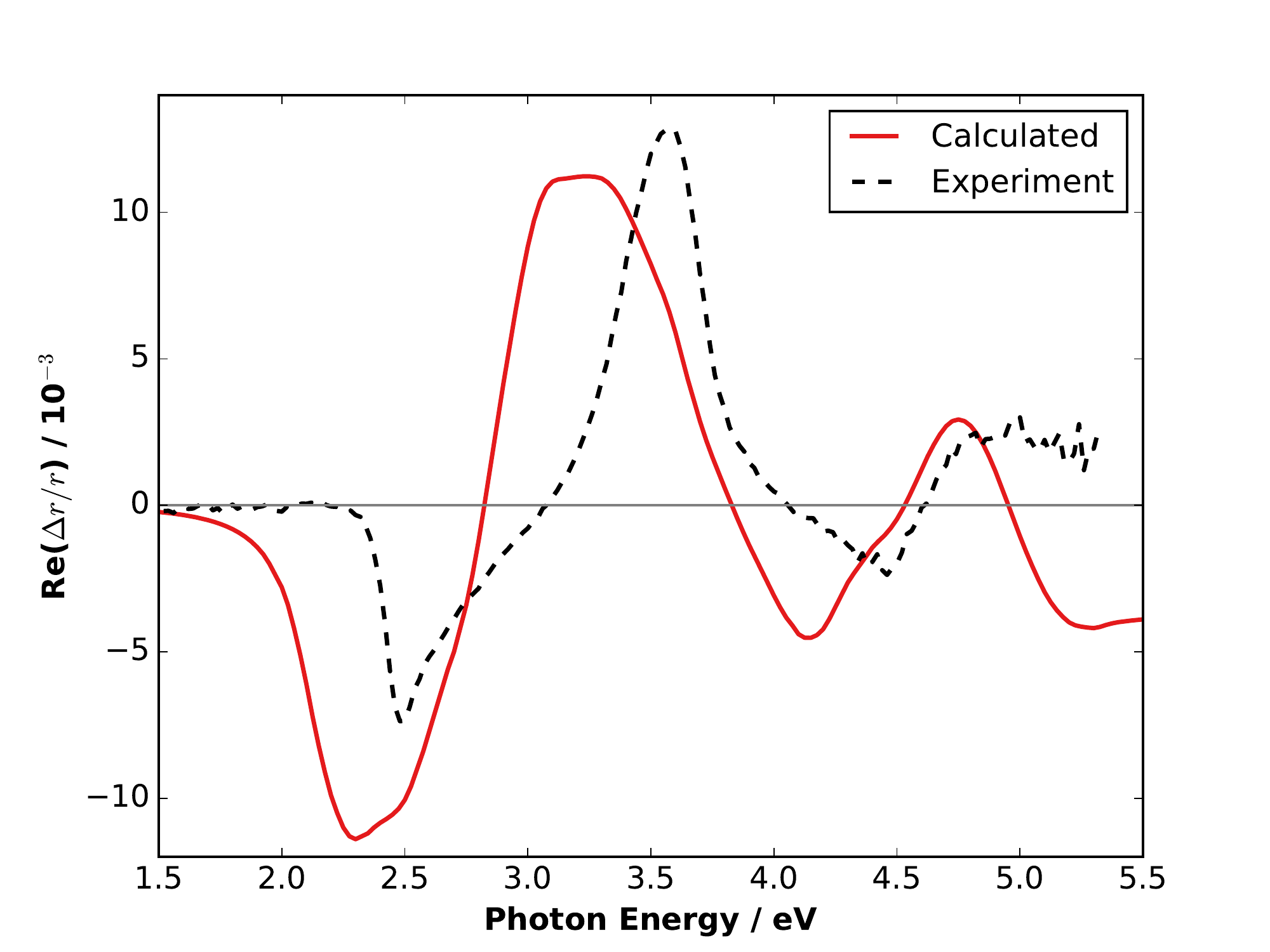}
 \caption{Calculated (red) and experimental (black, dashed)\cite{May_GaP_H2O_2013} RA spectrum at 300\,K of the pristine P-rich, $(2\times2)$-2D-2H surface.}
 \label{fig:P-rich}
\end{figure}

We first tested if the selected level of theory for the calculation of RAS is valid for the treatment of P-rich GaP(100) by comparing the computed spectrum with the experimental spectrum of the pristine surface. Figure~\ref{fig:P-rich} shows the experimental RA spectrum from Ref.\cite{May_GaP_H2O_2013} at 300\,K together with the spectrum calculated in YAMBO. We observe that the agreement is very good, both quantitatively and qualitatively. Compared to the calculation of Hahn et al.\cite{Hahn_P-rich_GaP_calculated_2003}, our spectrum shows slightly more narrow features, which can be attributed to the reduced broadening. Consequently, our chosen approach appears to be appropriate for GaP.

The experimental boundary condition that no oxygen is found on the surface, leaves us now with the ingredients Ga, P, and H. Under the assumption that the heavy metal cations are unlikely to evaporate at room temperature, there are in principle three options to modify the surface, the addition of hydrogen, the removal of hydrogen, or the removal of phosphorous (and hydrogen) in the form of e.g. phosphine. The removal of phosphorous does, however, appear very unlikely due to the experimental finding that already very gentle annealing, without resupply of phosphorous, can completely recover the surface.\cite{May_GaP_H2O_2013} For geometry optimisation, we relaxed the condition of the $c(2\times 2)$ surface symmetry, also considering additional structures.

To test if surface geometries are energetically favourable, adsorption energies, $E_{ads}$, were calculated. Here, $E_{ads}$ is the difference between the total energies of the slab with adsorbate, $E_{slab,ads}$ and the sum of the total energies of the slab without adsorbate, $E_{slab}$, and the (gas-phase) adsorbate, $E_{adsorbate}$:

\begin{equation}
\label{eq:ads_energy}
 E_{ads} = E_{slab,ads} - (E_{slab} + E_{adsorbate})
 \end{equation}

The resulting energies from geometry optimisation, which also include the formation of the surface reconstruction, with respect to a $(2\times1)$ unit cell are listed in Table~\ref{tab:ads_energies}. Our point of reference is the $(2\times1)$-1D structure, i.e. the P-dimer without hydrogen. An increase of the hydrogen coverage to the $(2\times2)$-2D-2H surface and then to the $(2\times2)$-1D-2P-4H surface decreases the energy with an increase of H-coverage and is therefore energetically beneficial. The lowest energy configuration is found for an even higher coverage, the $c(2\times2)$-2P-3H surface. A further increase of the hydrogen coverage to two H per P then increases the energy again. These adsorption energies from total energy considerations are, however, only a first order approximation. For higher accuracy of up to 0.1\,eV, the grand-canonical potential $\Omega$ with respect to the chemical potentials of the involved species has to be considered.\cite{Schmidt_InP_hydrogen_stabilized_2003} Furthermore, the surface structure can be dominated by kinetics, as it is often the case for experimental growth conditions at elevated temperatures.\cite{Brueckner_layer_removal_Si_2013, Romanyuk_interface_states_2016} The $c(2\times2)$-2P-3H surface structure was indeed already proposed for the closely related P-rich InP(100) surface under room-temperature atomic hydrogen exposure, where kinetics can be expected to be less dominant.\cite{Letzig_2x2_reconstruction_InP_2006}

\begin{table}[h]
\caption{Adsorption energies of different surfaces per $(2\times1)$ unit cell, calculated after eq.~\ref{eq:ads_energy}.}
\begin{center}
\begin{tabular}{l|l}
Surface & $\mathbf{E_{ads}}$\textbf{ / eV}\\
\hline
$(2\times1)$-1D & 0\\
$(2\times2)$-2D-2H & -0.96\\
$(2\times2)$-1D-2P-4H & -1.17\\
$c(2\times2)$-2P-3H & -1.55\\
$c(4\times4)$-2D-4P-8H & -1.00
\end{tabular}
\end{center}
\label{tab:ads_energies}
\end{table}

\begin{figure}[htb]
 \centering
 \includegraphics[width=\textwidth]{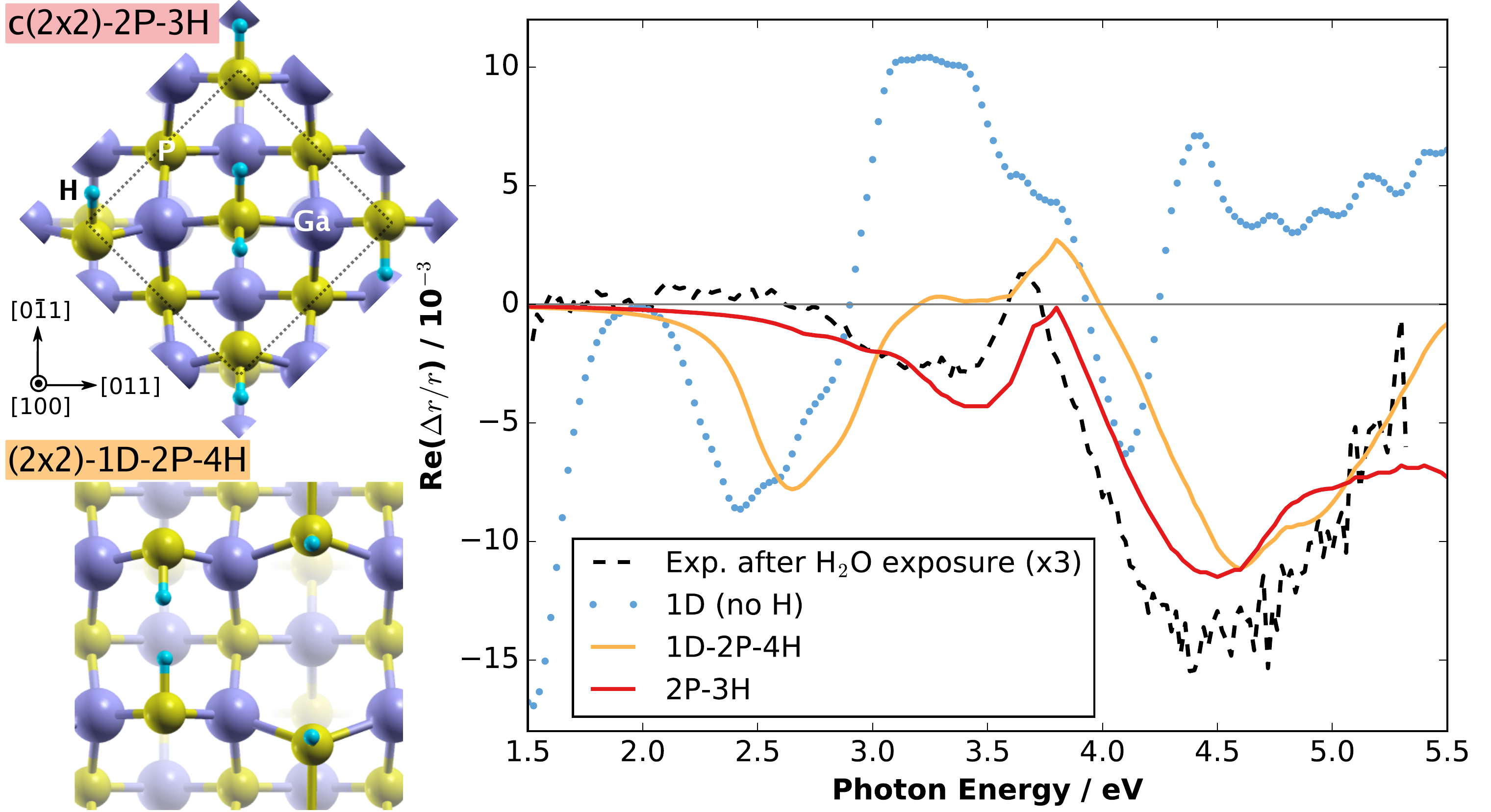}
 \caption{Structures of surfaces with different hydrogen coverages together with their respective RA spectra. The experimental spectrum (black, dashed curve) was scaled by a factor of 3.}
 \label{fig:H-rich-RAS-struct}
\end{figure}

A comparison of calculated RA spectra for the different surface geometries is shown in Fig.~\ref{fig:H-rich-RAS-struct}. A notable feature of the surface structure without hydrogen (blue dotted curve) is the strongly negative peak at 1.5\,eV. This low-energetic transition is owed to a reduced bandgap. The rest of the optical anisotropy is rather similar to the calculation for the $(2 \times 2)$-2D-2H surface (Fig.~\ref{fig:P-rich}). The spectrum for the $(2\times2)$-1D-2P-4H surface (orange line) resembles the experimental spectrum after H$_2$O exposure (dashed curve) already to a higher extent, especially in the range beyond 3.8\,eV. The characteristic negative anisotropy around 2.5\,eV, which does not arise in the experimental spectrum of the exposed surface, still exists, though shifted to higher energies. Finally, the calculated spectrum for the $c(2\times2)$-2P-3H is in very good agreement with experiment, with the characteristic broad negative peak between 3 and 3.5\,eV, and a very strong negative feature beyond 4\,eV. Energetic discrepancies between the experimental and calculated spectra are expected due to the bandgap errors of PBE.

Due to the quantitative nature of RAS, a reduced intensity -- as observed for the experimental spectrum -- indicates a surface that is not completely ordered or that exhibits mutually perpendicular surface domains, which cancel to zero. The surface slab for the calculation is, however, perfectly ordered, with all of the surface in terminated by the same reconstruction. This consequently leads to higher intensities. The $c(2\times2)$-2P-3H appears to be the most likely structure of the P-rich GaP(100) surface after exposure to gaseous water, both from an energetic and theoretical spectroscopy perspective. Quantitatively, however, the spectroscopy suggests that some disorder persists.

While the experimental averaging of RAS is too extensive both in time and (in-plane) space to deduce the reaction path, the timescale for DFTMD is too short. Therefore, we can only try to deduce the reaction path from the final state of the surface. A single-domain, $c(2\times2)$-2P-3H surface would exhibit a three times higher hydrogen coverage than the initial $(2 \times 2)$-2D-2H surface. In principle, this hydrogen could stem from dissociated water molecules, but as no oxygen was found on the surface after exposure, this would require a full oxygen evolution reaction (with desorption of molecular O$_2$), which has a very high energy barrier. Though the reaction could in principle be facilitated by photoexcitation of the semiconductor from ambient illumination or probe light of the spectrometer, we consider this reaction unlikely to take place at 100\% Faradaic efficiency. If we do, however, also consider the reduced signal intensity of the experimental spectrum, another pathway becomes plausible: The water molecules could induce a hydrogen-hopping on the surface, leading to surface domains with hydrogen-accumulation, and other surface domains with hydrogen depletion. If the total hydrogen coverage is conserved, the ratio of these domains will be 1:2 and if we assume that the hydrogen-depleted domains are less ordered, i.e. optically isotropic, this explains the threefold increased intensity of the calculated spectrum.

An increased hydrogen -- or possibly proton -- mobility on the surface induced by water would actually be a very beneficial behaviour for solar water splitting applications, facilitating the combination of two adsorbed H atoms to molecular hydrogen, the Tafel step. The extraordinary stability of the P-rich surface against oxidation by water is, however, a unique feature of GaP(100), the closely related P-rich InP(100) does not show this behaviour.\cite{May_InP_H2O_2014} A P-rich termination of a GaInP solar cell for water splitting might therefore increase the catalytic activity of the semiconductor for the hydrogen evolution reaction, but might require a stoichiometry with more Ga than In. If the P-rich surface also remains stable in bulk (as opposed to gas-phase) water, remains to be seen.

\begin{figure}[h!]
 \centering
 \includegraphics[width=\textwidth]{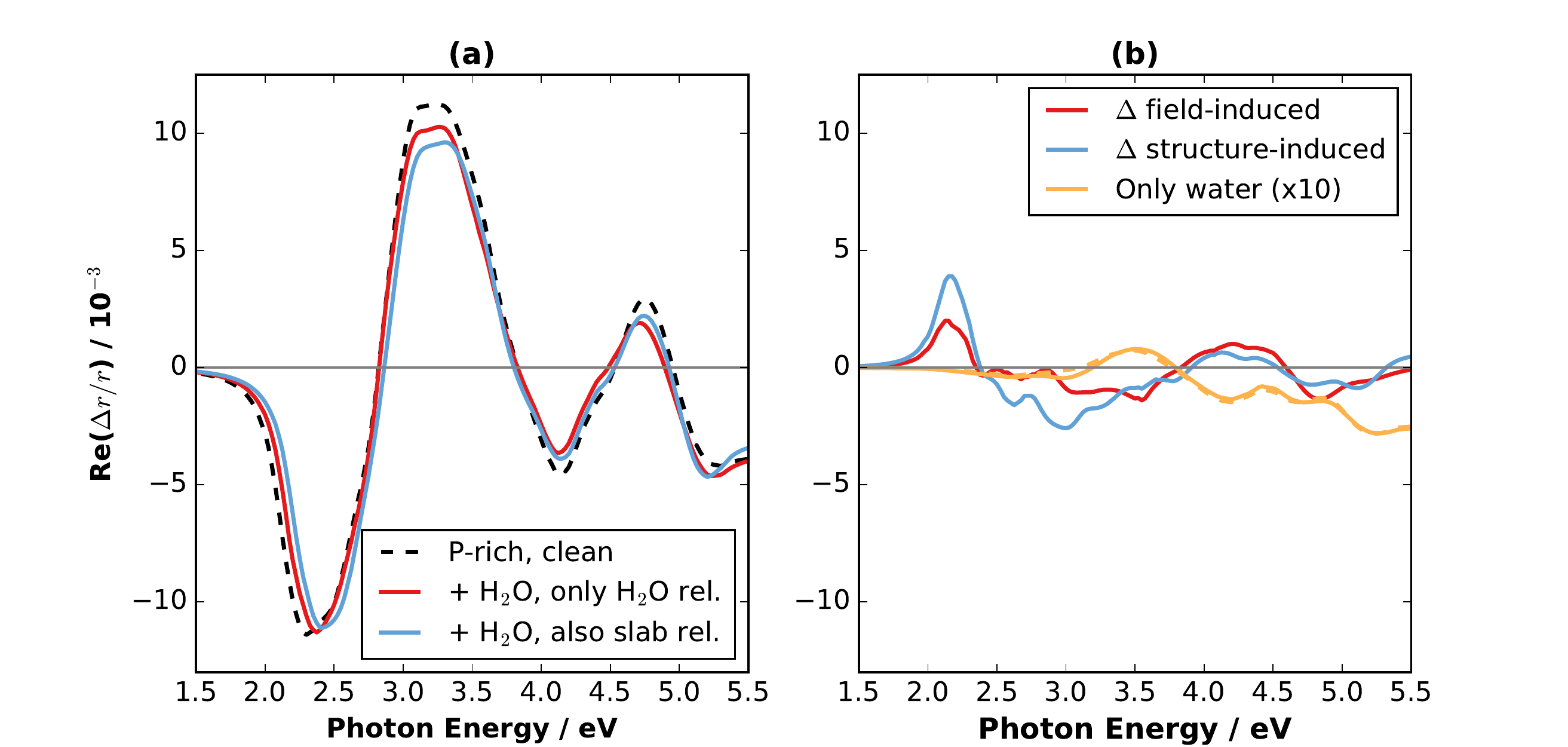}
 \caption{(a) Calculated RAS for the P-rich surface without water (dashed line) and below 2\,ML of water with structural relaxation of only the water (red) and also the slab (blue). (b) Differences to the surface without water. The spectrum for water only (orange; solid line for relaxation limited to water only) was scaled by a factor of 10.}
 \label{fig:RAS-water}
\end{figure}

To test the feasibility of theoretical RA spectroscopy of GaP(100) surfaces in contact with H$_2$O towards electrochemical investigations, we modelled the P-rich, $(2\times2)$-2D-2H surface in contact with 2 monolayers (ML) of water. Spectra were calculated for the full half-slab (Fig.~\ref{fig:RAS-water}a), either for a relaxation of water and surface (solid blue line) or only water (solid red line). The contribution of water can be separated out by the real-space cutoff and is with a maximum of $<3\cdot 10^{-4}$ in the energy range upt to 6\,eV almost negligible (orange curves in Fig.~\ref{fig:RAS-water}b) due to the large bandgap. The bandgap of pure water is with 4.5\,eV largely underestimated on the GGA level employed here.\cite{Opalka_ionization_potential_hydroxide_2014} However, the hereby induced early onset of optical transitions in water is alleviated by the typical limitation of the experimentally accessible energy range to below 6\,eV.

The impact of the water molecules on the spectra of the GaP surface itself is an indirect one as evidenced by a comparison of two relaxation approaches during geometry optimisation: When both water and surface are allowed to relax, the difference to the spectrum without water is significant, showing both an energetic blue-shift of the low-energy peak around 2.4\,eV, a reduction of the positive signal at 3.3\,eV, and minor changes beyond 4\,eV.

The spectrum, where the GaP surface atoms are fixed during the geometry optimisation, shows an intermediate behaviour with a subdued blue-shift in the low-energetic region, a small red-shift of the positive signal beyond 4.5\,eV, and an intermediate reduction of signal intensities. The difference spectra (Fig.~\ref{fig:RAS-water}b) are with 2 to 4$\cdot10^{-3}$ units rather intense, especially in the range around 2.2, 3, and 4.2\,eV. This would make time-resolved \textit{in situ} experiments feasible, where a high time-resolution is obtained by fixing the photon energy to probe for instance the impact of an applied potential.

The signal change for the fully relaxed system stems from both, water double layer effects that include the linear electro-optic effect from electric fields,\cite{Leo_PRB_1989} and water-induced structural changes of the GaP surface. For the partially relaxed surface, there is no structural effect and as the water does not directly contribute to the optical anisotropy, the spectral modification arises from the electric field induced by the water that changes the dielectric function of the GaP surface. These observations emphasize the potential of RAS at the semiconductor--electrolyte interface, where both surface-ordering and surface electric fields can be probed in the electrolyte. The computational effort for modelling, that is required for the understanding of experimental features, is alleviated by the fact that water itself does not contribute significantly to the spectrum. This removes the necessity to average many DFTMD snapshots of bulk water molecules. Averaging over structural fluctuations of the surface itself is, however, still required, with the extent depending on the rigidity of the solid surface.

\section{Summary \& Outlook}

The comparison of experimental data with theoretical spectroscopy provides strong evidence for the formation of an H-rich, $c(2\times2)$-2P-3H GaP(100) surface phase after contact of the initially P-rich surface with gas-phase water. Our prediction that the surface undergoes a phase separation with optically isotropic regions and ordered $c(2\times2)$-2P-3H domains should be verifiable by further experiments, such as scanning tunnelling microscopy or high-resolution electron energy loss spectroscopy. The enhanced, water-induced hydrogen- or proton-mobility required for the surface transition could benefit the catalytic activity of a P-rich GaP(100) surface for a hydrogen evolution reaction,\cite{Wood_surface_chemistry_water_InP_GaP_2014} if the surface is also stable in an electrolyte.

Our results also show that RAS is a spectroscopic technique attractive for \textit{in situ} studies of semiconductor--liquid interfaces. Spectra are not only sensitive to structural changes, but also to water-induced electric fields that play an important role in the Helmholtz-layer, rendering RAS an indirect probe for interface energetics. The effect of water itself on the optical anisotropy in the energetic region typically probed by a spectrometer is comparatively small, which greatly alleviates modelling, as the necessity to average many CPMD snapshots of bulk water is lifted. Further work will investigate the impact of charged surfaces in a bulk electrolyte on spectral features and interface energetics.

\ack

Computational resources were provided UK Car-Parrinello (UKCP) consortium funded by the Engineering and Physical Sciences Research Council (EPSRC) of the United Kingdom as well as Helmholtz-Zentrum Berlin für Materialien und Energie. MMM acknowledges funding from the fellowship programme of the German National Academy of Sciences Leopoldina, grant LPDS 2015-09.

\section*{References}
%\bibliographystyle{iopart-num}
%\bibliography{Literatur.bib}
\providecommand{\newblock}{}

\end{document}